\documentclass[10pt,aps, prl,reprint,superscriptaddress, showpacs]{revtex4-1}
\usepackage[utf8x]{inputenc}
\usepackage[pdftex]{graphicx}
\usepackage{amsmath,amssymb}
\newcommand{\trace}[2]{\text{Tr}_{#1}\left[#2\right]}
\newcommand{\fig}[1]{Fig.~\ref{#1}}

\newcommand{\nn}{\nonumber}

\usepackage{color}

\begin{document}

\title{Distribution of energy dissipated by a driven two-level system}

\author{Philip Wollfarth}
\affiliation{Institut f\"ur Theorie der Kondensierten Materie, Karlsruher Institut f\"ur Technologie, 76128 Karlsruhe, Germany}
\affiliation{DFG-Center for Functional Nanostructures (CFN), 76128 Karlsruhe, Germany}

\author{Alexander Shnirman} 
\affiliation{Institut f\"ur Theorie der Kondensierten Materie, Karlsruher Institut f\"ur Technologie, 76128 Karlsruhe, Germany}
\affiliation{DFG-Center for Functional Nanostructures (CFN), 76128 Karlsruhe, Germany}

\author{Yasuhiro Utsumi} 
\affiliation{Department of Physics Engineering, Faculty of Engineering, Mie University, Tsu, Mie, 514-8507, Japan}

\begin{abstract}
In the context of fluctuation relations, we study the distribution of energy dissipated by a driven two-level system. 
Incorporating an energy counting field into the well known spin-boson model enables us to calculate the distribution function of the amount of energy exchanged between the system and the bath. We also derive the conditional distribution functions of the energy exchanged with the bath for particular initial and/or final states of the two-level system. 
We confirm the symmetry of the conditional distribution function expected from the theory of fluctuation relations. 
We also find that the conditional distribution functions acquire considerable quantum corrections at times shorter or of the 
order of the dephasing time. Our findings can be tested using solid-state qubits. 
\end{abstract}


\date{\today}

\maketitle


After the discovery of universal relations out of equilibrium, i.e. the fluctuation relations (FRs), e.g., Crooks and Jarzynski relations \cite{bochkov-1987,Campisi-2011, Esposito-2009, Seifert-2012}, 
it has been recognized that fluctuations of the entropy (or the heat and work) and micro-reversibility are the key concepts 
relevant for the dynamics far from equilibrium. 
The impact of entropy fluctuations becomes pronounced as the system size decreases. 
Thus, the FRs have been tested at room temperature in relatively small systems, such as colloidal particles and biomolecules \cite{Wang-2002, Liphardt-2002, Collin-2005}. 
During the last few years, by using quantum dots, the FRs have been demonstrated at the single electron level at temperature as low as 100 mK \cite{Utsumi-2010, Kueng-2012, Saira-2012, Koski-2013}. 
Almost all observed results are well explained within a classical 
stochastic picture, in which individual random trajectories of the system are well defined~\cite{Seifert-2012}.

Recently, several attempts have been made~\cite{Nakamura-2010, Albash-2013, Batalhao-2013} towards a generalization into the quantum regime, where it would be difficult to define work unambiguously. 
An early experiment~\cite{Nakamura-2010} used an Aharonov-Bohm interferometer to test the average and the variance of the electric current probability distribution function (PDF)~\cite{Saito-2008, Foerster-2008, Andrieux-2009}. 
Recent discussions \cite{Batalhao-2013, Campisi-2013, Dorner-2013, Mazzola-2013} are focused on the Crooks FR \cite{Crooks-1999} for a driven qubit. The Crooks FR relies on the concept of work performed along each individual trajectory~\cite{Collin-2005}. However, for quantum systems, there is a fundamental problem to define this work~\cite{Talkner-2007,Campisi-2011}. This problem has motivated research toward FR in quantum systems. 
In a recent experiment~\cite{Batalhao-2013}  the characteristic function (CF), i.e., 
the Fourier transform of the energy PDF, was measured by the Ramsey interferometry of an ancillary qubit \cite{Campisi-2013, Dorner-2013, Mazzola-2013}. This approach relies on the correspondence between the Loschmidt echo and the CF \cite{Silva-2008, lesovik-2006}. A straightforward approach based on the measurement of thermodynamic 
quantities, e.g., energy would be still desirable. 

In this letter, we are motivated by the idea of calorimetric measurements of the energy $\epsilon$ dissipated into the heat bath coupled to a quantum system \cite{Pekola-2013, Hekking-2013}. 
This approach can be realized in a superconducting qubit coupled to a resistor, whose temperature 
is monitored in a time-resolved fashion~\cite{Timofeev-2013}. 
We further propose to pre- and post-measure the state of the qubit and calculate the 
conditional probabilities to dissipate energy $\epsilon$ given the initial and the final state.

When the qubit and the bath are not coupled, 
the FR trivially ensures that the transition probability from the initial state to the final state under the forward driving ${\mathcal P}_\tau(f|i)$ is equal to that of time reversal process under the backward driving ${\mathcal P}_{\tau,B}(i|f)$. 
If the qubit is coupled to the bath, the probability density of energy exchanged between the qubit and bath with particular initial and final states ${\mathcal P}_\tau(\epsilon,f|i)$ can be defined and the detailed fluctuation relation by Jarzynski \cite{Jarzynski-2000}, eq. \eqref{eq:FR}, in quantum regime can be checked. Our result can be seen as an extension to recent results \cite{Gasparinetti-2014}, where the distribution of the total energy absorbed by the bath has been calculated. 
In addition, we find that due to the the final state selection, off diagonal elements of the density matrix provide an important correction to the conditional probability distributions. These quantum corrections emerge solely for the pre- and post-selected 
distributions.

A general Hamiltonian of a periodically driven dissipative system reads
\begin{align}
 H(t) = H_S(t) + H_I + H_B\ .
\end{align}
Here $H_S(t)=H_S(t+\tau_p)$ is the periodically driven system's Hamiltonian with period $\tau_p=2\pi/\omega$. The coupling is given by $H_I =  \sum_\alpha A_\alpha \otimes B_\alpha$, where $A_\alpha, B_\alpha$ represent operators of the system and the bath respectively. The bath's Hamiltonian can be choses as $H_B = \sum_\alpha \omega^{\phantom\dagger}_\alpha b^\dagger_\alpha b^{\phantom\dagger}_\alpha$. 

Using the well established Floquet theory~\cite{Grifoni-1998} the Hamiltonian $H_S(t)$ can be made time-independent
by introducing the extra quantum number corresponding to the number of quanta of the driving field absorbed by the system. 
In this case, naturally, the complexity of the system bath interaction $H_I$ rises.  
In this paper we consider systems, in which the full-fledge Floquet technique can be replaced by a simpler scheme, where the time dependency of $H_S(t)$ can be eliminated by a transformation in the rotating frame. This is achieved by a time dependent rotation matrix $R(t)$ such that the resulting system Hamiltonian $\widetilde{H}_S(t)= R(t)H_S(t)R^\dagger(t) +i \dot{R}(t)R^\dagger(t)$ becomes time-independent. The total Hamiltonian in the rotating frame reads then
%
%
%
\begin{align}
\label{HRotated}
 \widetilde{H} = \widetilde{H}_S + \widetilde{H}_I(t) +H_B\ ,
\end{align}
where the periodic time dependency has been shifted to the interaction part $\widetilde{H}_I(t) = \sum_\alpha R(t)A_\alpha R^\dagger(t) \otimes B_\alpha$. 

With this preliminary, we calculate the conditional PDF of energy $\epsilon$ dissipated to 
the bath with the initial and the final state selection:
\begin{align}
 \mathcal{P}_\tau(\epsilon,f|i) = \sum_n \sum_k\sum_{\sigma=-1,0,1} p^{k,n,\sigma}_\tau( f|i) \delta(\epsilon - n\omega - \sigma \Omega_k).
\end{align}
The energy is quantized to multiples of the driving frequency $\omega$ plus the level spacings $\Omega_k$ of the system in 
the rotating frame. The PDF is determined by the weights $p^{k,n,\sigma}_\tau(f|i)$. 
The indices $i,f$ indicate the initial and final state selection and the normalization condition reads $\int d\epsilon \sum_f \mathcal{P}_{\tau}(\epsilon,f|i)=1$.
The calculation of the PDF is performed via the characteristic function(CF)
$\chi_\tau(\lambda,f|i) = \int d\epsilon \,e^{i\epsilon \lambda} \mathcal{P}_\tau(\epsilon,f|i)$ which we calculate using the method of full counting statistics (FCS)~ \cite{Esposito-2009,Wollfarth-2013}.
We obtain 
\begin{align}\label{eq:GF}
 \chi_\tau(\lambda,f|i) = \trace{}{|f\rangle \langle f|e^{\mathcal{L}(\lambda) \tau}\rho_i(\lambda,0)},
\end{align}
with $\mathcal{L}(\lambda)$ being the Liouvillian super-operator modified  by inclusion of the counting field whereas 
$\rho_i(\lambda,0)$ the initial density matrix. The projector $|f\rangle \langle f|$ is responsible for the post-selection of the desired final state. Following Ref.~\onlinecite{Esposito-2009}
we build in the counting field into the Hamiltonian $H(\lambda) \equiv e^{iH_B \lambda/2} H e^{-iH_B\lambda/2}$, where the bath Hamiltonian $H_B$ or more specifically the energy emitted / absorbed by the bath is the quantity which we want to count. 
As the bath Hamiltonian commutes with everything except the interaction term, we obtain a modified interaction part $H_I(t)(\lambda) = \sum_\alpha \widetilde{A}_\alpha(t) \otimes e^{iH_B\lambda/2}B_\alpha e^{-iH_B\lambda/2}$.

In this paper we focus on the case of a driven two-level system. 
For the derivation of the CF we use a master equation approach similar to \cite{Breuer, Esposito-2009}. Our starting point is a Markovian master equation in the interaction picture, where we assume the total density matrix being initially factorized $\rho(0) = \rho_S(0)\otimes \rho_B$. The indices $S$ and $B$ denote here system and bath respectively.  
Within secular approximation we obtain the following master equation
\begin{align}
 \frac{d}{dt} \vec{\rho}(t) = M(\lambda) \vec{\rho}(t)\ ,
\end{align}
where 
\begin{align}\label{eq:S_O}
 M(\lambda) =\left(\begin{array}{cccc}
                    -\Gamma_{gg} & \Gamma_{ge}&0&0\\
                    \Gamma_{eg}&-\Gamma_{ee}&0&0\\
                    0&0&+i\Omega-\Gamma_\varphi&0\\
                    0&0&0&-i\Omega-\Gamma_\varphi
                   \end{array}
\right)
\end{align}
is a super-operator containing the transition rates $\Gamma_{ij}$ and the dephasing rates $\Gamma_\varphi$.
The secular approximation, which amounts to neglecting all other dissipative rates in (\ref{eq:S_O}), is well justified 
provided $\Omega \gg \Gamma_\varphi,\Gamma_{ij}$. The rates $\Gamma_\varphi(\lambda)$ and $\Gamma_{ij}(\lambda)$ depend on the counting field $\lambda$. 

The reduced density matrix of the system is represented by a 4-component vector $\vec{\rho}(t)= \left(\rho_{gg}(t),\rho_{ee}(t),\rho_{eg}(t),\rho_{ge}(t)\right)^T$. Within this representation, the generating function for the conditional probabilities can be 
split to a classical part $\chi^c_\tau(\lambda, f|i)$, which solely depends on the populations $\rho_{gg}(t),\rho_{ee}(t)$, and 
to a quantum part $\delta\chi_\tau(\lambda,f|i)$, which contains the information about the coherences 
$\rho_{eg}(t),\rho_{ge}(t)$. The conditional PDF, thus, reads
\begin{eqnarray}
 &&\mathcal{P}_\tau(\epsilon, f|i) = \mathcal{P}^{c}_{\tau}(\epsilon, f|i) + \delta   \mathcal{P}_{\tau}(\epsilon, f|i)
 \nonumber\\
&&= \frac{1}{2\pi} \int d\lambda e^{-i \lambda \epsilon }  \left(\chi_\tau^c(\lambda, f|i) + \delta \chi_\tau(\lambda, f|i)\right) \ .
\end{eqnarray}
One can easily observe from Eq.~\eqref{eq:GF} that the quantum contributions cancel each other in the total (unconditional) 
PDF of energy dissipated to the bath, $\mathcal{P}_\tau(\epsilon,i)=\sum_f \mathcal{P}_\tau(\epsilon,f|i)$.

As an example, we analyze a two-level system driven by a circularly polarized field. The Hamiltonian reads $H_S(t)=-\frac{\omega_0}{2}\sigma_z + \frac{\Omega_R}{2}\left(\cos(\omega t) \sigma_x - \sin(\omega t)\sigma_y\right)$, where $\omega_0$ denotes the level splitting in the laboratory frame and $\Omega_R$ is the Rabi-frequency. We further set $\hbar=1$. The 
transformation to the rotating frame, discussed above, is provided in this case by $R(t)=\exp (-i\sigma_z \omega t/2)$ and the resulting Hamiltonian in the rotating frame reads $\widetilde{H}_S = -\frac{\Delta}{2}\sigma_z +\frac{\Omega_R}{2} \sigma_x$, 
where we introduced the detuning $\Delta = \omega_0-\omega$. By applying second rotation  $R_2= \exp(-i\sigma_y\theta/2)$ with $\tan \theta = \Omega_R/\Delta$ the system is diagonalized into its energy eigenbasis $\{|g\rangle, |e\rangle\}$.


\begin{figure}[h!!!!!!!!!!!!!!!!!]
\begin{center}
\includegraphics[width=\linewidth]{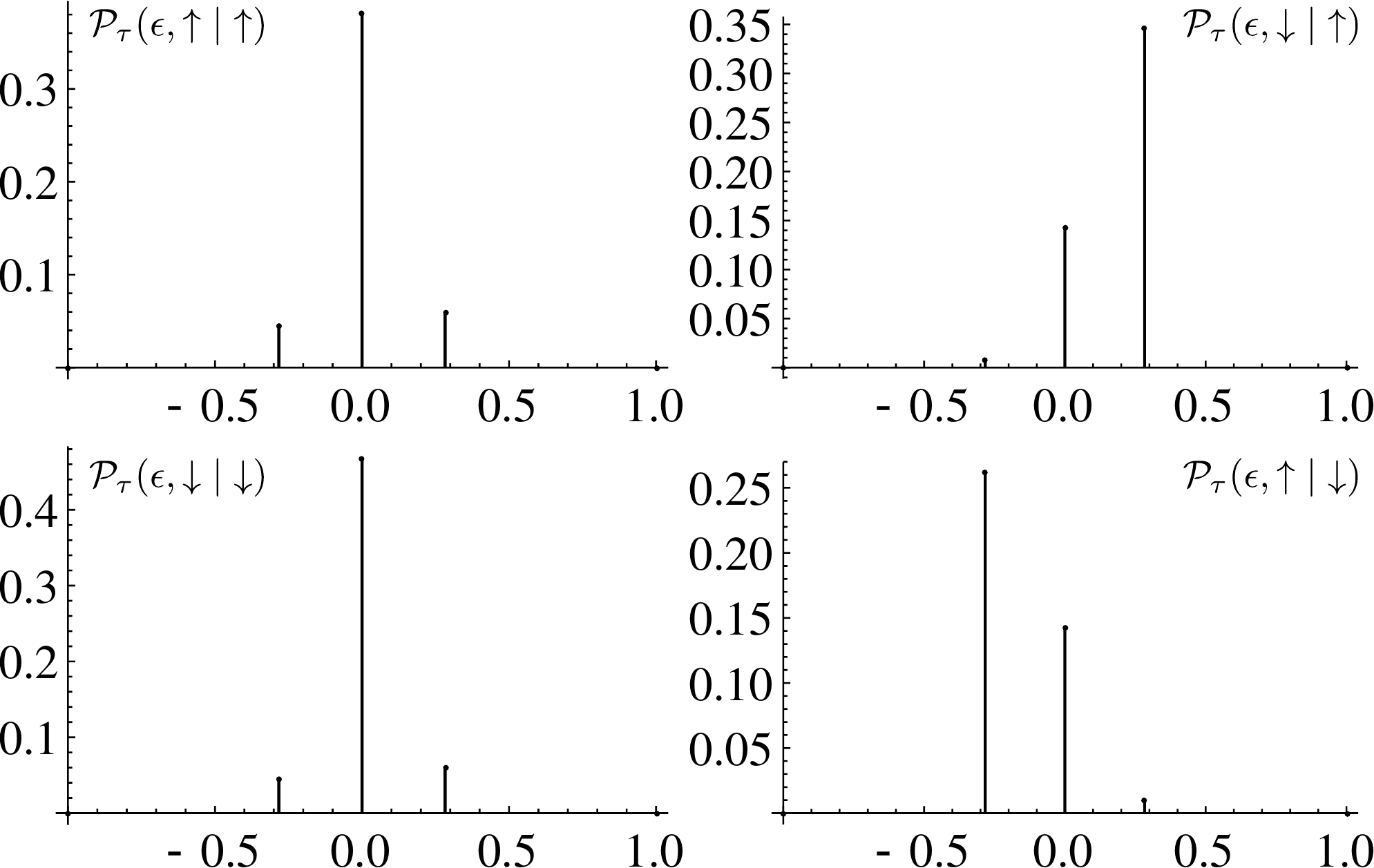}
\caption{Conditional probability densities $\mathcal{P}_\tau(\epsilon,f|i)$ of energy emitted to the bath. Here, $f (i)$ indicates the final (initial) state selection in the rotating frame. The energy $\epsilon$ is normalized to the driving frequency $\omega$. The probability densities are plotted for coupling $\gamma_0 = 0.01 \omega$, temperature $\beta = 1/ \omega$, detuning $\Delta=0.2 \omega$, Rabi-frequency $\Omega_R = 0.2 \omega$ and a driving time $\tau =200/\omega$. We used Gaussian smearing and rescaled the peaks accordingly to visualize the weighted delta peaks in the PDF. }
\label{pic:pzall}
\end{center}
\end{figure}
We consider, first the case of longitudinal coupling to the bath, i.e., $A_i= \sigma_z$. 
In this case the transformation $R(t)$ does not modify the interaction Hamiltonian in  
(\ref{HRotated}), i.e., $\tilde H_I = H_I$. To get a better insight into the problem, we use the previously mentioned Floquet-picture. We consider the driving terms $e^{\pm i \omega t} \equiv e^{\pm i \phi}$ as raising and lowering operators of energy quanta $\omega$ absorbed and emitted by the bath. As the interaction term remains invariant under the rotation $R(t)$, no transitions between different Floquet copies of the system occur. In other words, the energy quanta $\omega$ cannot be exchanged between the system and the bath. Thus, the only dissipative processes that can occur are those where the energy of the level splitting in the rotating frame $\Omega=\sqrt{\Delta^2 +\Omega_R^2}$ can be exchanged between the system and the bath. 
The calculation is performed for a bath characterized by the correlation function $\gamma(\omega) \equiv \int_{-\infty}^\infty ds \langle B^\dagger (s)B(0)\rangle e^{i\omega s} = \gamma_0 \omega/(1-e^{-\beta \omega})$, where $\gamma_0$ denotes the coupling strength and $\beta= (k_BT)^{-1}$. 
The results for the conditional probability densities for the energy emitted to the bath are depicted in \fig{pic:pzall}. 

\begin{figure}[h!!!!!!!!!!!!!!!!!]
\begin{center}
\includegraphics[width=\linewidth]{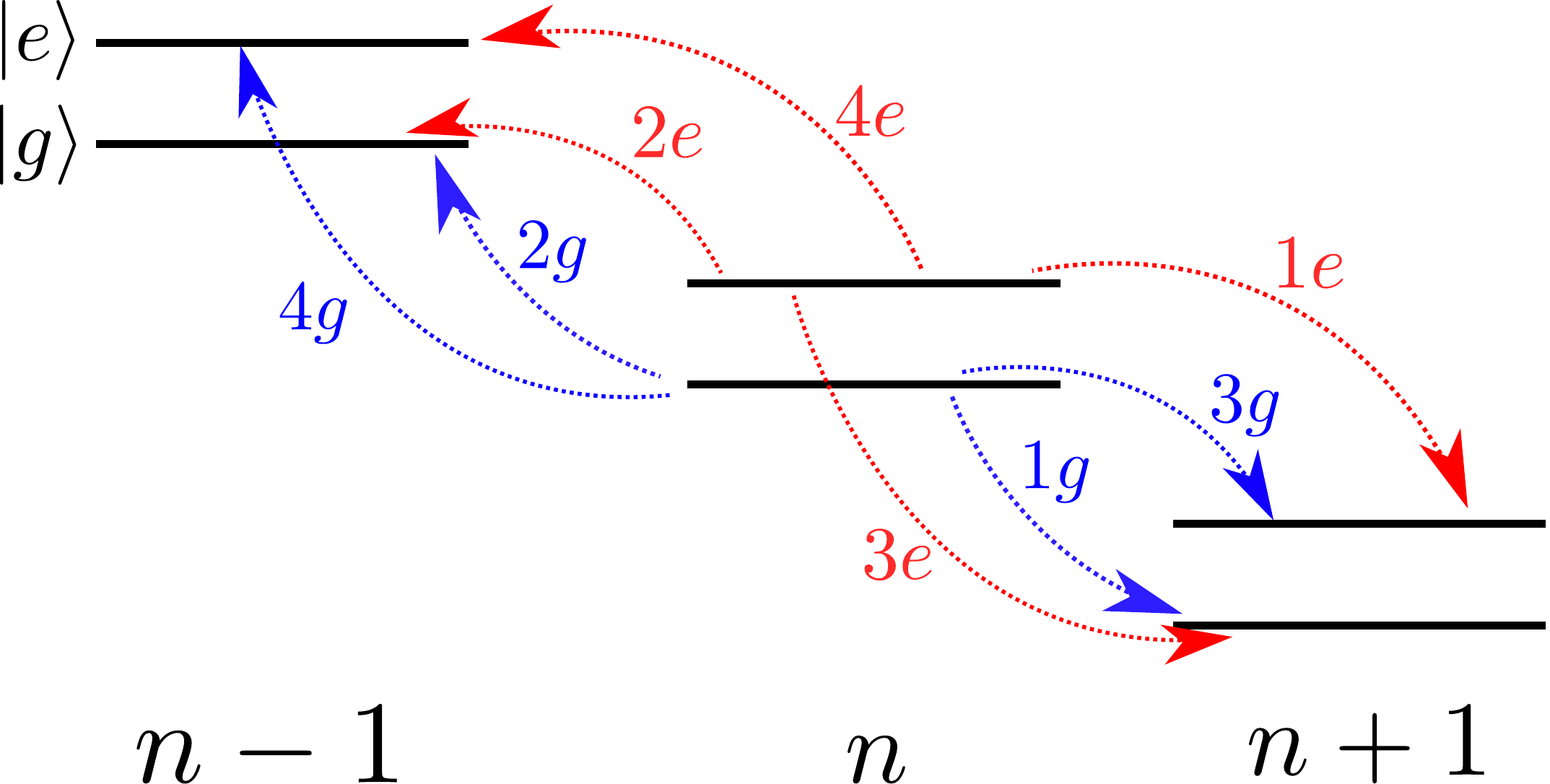}
\caption{(Color online) Floquet-picture of the two level system in the energy-eigenbasis in the case of transversal coupling. The index $n$ depicts the $n$-th Floquet-copy of the system, which is energetically shifted to the system by the Frequency $n\, \omega$. There are 8 possible transitions. If the system was in the ground state $|g\rangle$ of the rotated frame, then only (blue) transitions with energy exchange of $\pm \omega$ ($1g,2g$) or $\pm\omega -\Omega$ ($3g,4g$) are possible. If the system has been in it's excited state $|e\rangle$, only the (red) transitions ($1e-4e$) are possible.}
\label{pic:floquet}
\end{center}
\end{figure}

A more interesting situation occurs in the case of transversal coupling $A_i=\sigma_x$. Here, the rotation matrix $R(t)$ does not commute with $H_I$ and, therefore, the coupling in the rotating frame reads $\widetilde{H}_I = \sum_i (e^{i\omega t} \sigma_+ + e^{-i\omega t} \sigma_-) \otimes B_i$.
In this case energy quanta of $\pm\omega$ can be exchanged between the system and the bath. The possible transitions are depicted in \fig{pic:floquet}. It is easy to see that the available transition frequencies 
depend on the current state of the system. Being in the ground state $|g\rangle$ of the rotating frame 
the system can make transitions with frequencies $\pm \omega$ to the neighboring Floquet copies of $|g\rangle$ or 
transitions with frequencies $\pm \omega - \Omega$ to the neighboring Floquet copies of $|e\rangle$. The transitions 
with frequencies $\pm\omega +\Omega$ are blocked. If the system is in the excited state $|e\rangle$, the transitions 
with frequencies $\pm \omega - \Omega$ are blocked. This explained why the conditional PDF's depend on the 
initial state of the system. 

The rates determining the evolution of the diagonal elements of the density matrix read

\begin{align}
 \Gamma_{gg}&= \gamma^-(-\Omega) - \frac{\sin^2\theta}{4} g(\omega,\lambda),\\
 \Gamma_{ge}&= \cos^4\frac{\theta}{2} \gamma(\Omega+\omega)e^{i\lambda (\Omega+\omega)}  +\sin^4\frac{\theta}{2} \gamma(\Omega-\omega)e^{i\lambda(\Omega-\omega)}\,\\
 \Gamma_{eg}&=\cos^4\frac{\theta}{2}\gamma(-\Omega-\omega) e^{i\lambda(-\Omega-\omega)}  \nn\\
 &+\sin^4\frac{\theta}{2} \gamma(-\Omega+\omega) e^{i\lambda(-\Omega+\omega)},\\
 \Gamma_{ee}&=\gamma^+(\Omega)- \frac{\sin^2\theta}{4} g(\omega,\lambda),
\end{align}
where
\begin{align}
 \gamma^\pm(\Omega ) &= \cos^4\frac{\theta}{2} \gamma(\Omega \pm \omega)   +  \sin^4\frac{\theta}{2} \gamma(\Omega \mp \omega),\\
 g(\omega,\lambda)&= \gamma(\omega)\!\left(e^{i\lambda\omega} -1\right) + \gamma(-\omega)\left( e^{-i\lambda\omega}-1\right).
\end{align}
For the dephasing rate we obtain

\begin{align}\label{Gphi}
 \Gamma_\varphi(\lambda)\!\!&= \!\!\frac{\sin^2\!\theta}{4}  \!\left(\gamma (\omega\!) \left(e^{i\omega \lambda}\!\! +1\right)+\! \gamma(-\omega\!) \left(e^{\!-i\omega \lambda}+1\right)\right) \nn\\
 &+\frac{1}{2}\left[\gamma^+(\Omega)+\gamma^-(-\Omega)\right].
\end{align}

\begin{figure}[h!!!!!!!!!!!!!!!!!]
\begin{center}
\includegraphics[width=\linewidth]{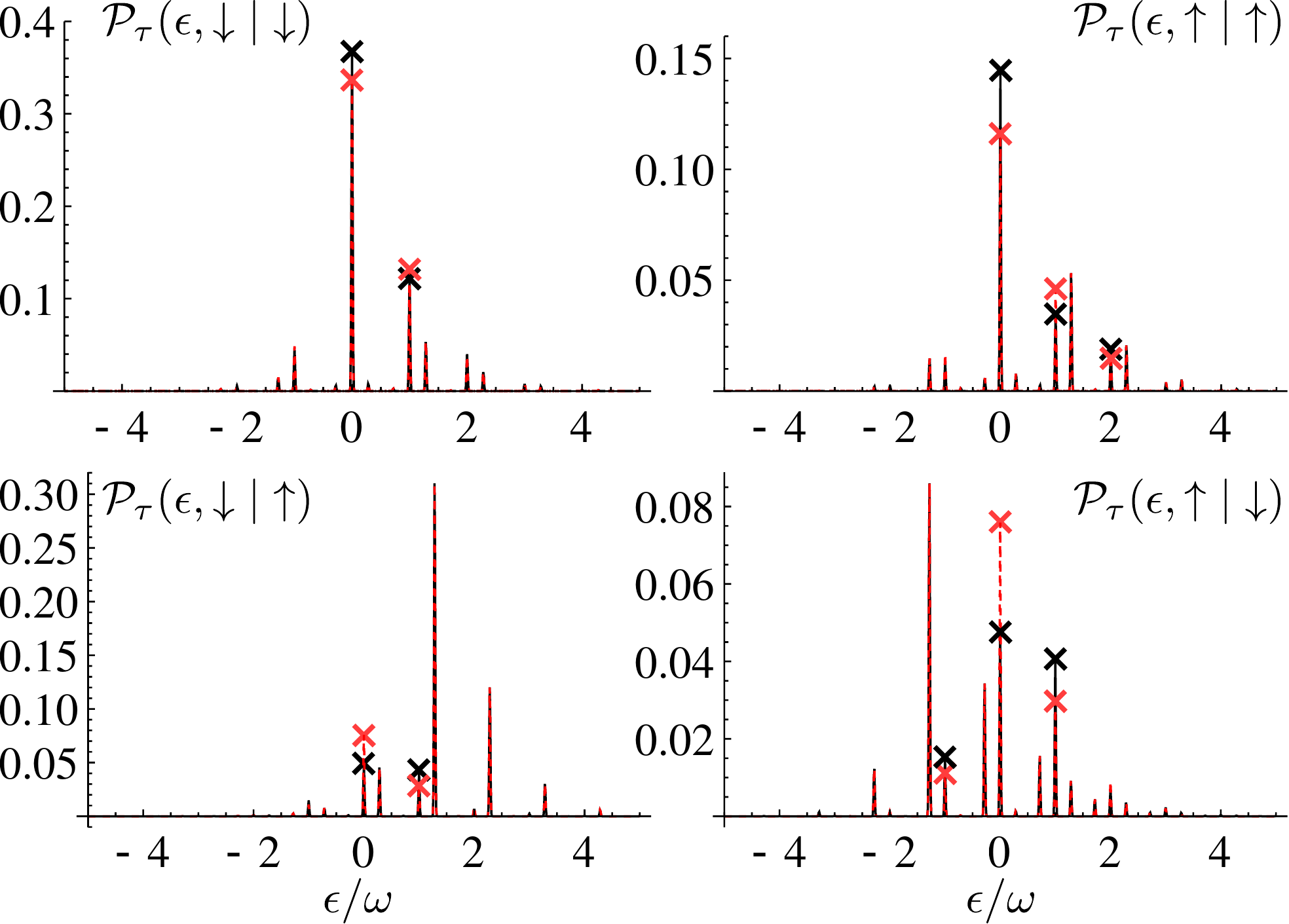}
\caption{ Conditional probability densities $\mathcal{P}_\tau(\epsilon,f|i)$ as a function of normalized energy $\epsilon/\omega$. The intervals of the peaks are given by the driving frequency $\omega$ or the level splitting $\Omega$. Plots are calculated for same parameters as for $\sigma_z$ coupling. The distinct choice of initial and final state drastically changes the structure of the probability densities. The markers denote the difference between the classical PDF with (black solid) and without (red dashed) quantum corrections $\delta\mathcal{P}_\tau(\epsilon,i|f)$ at integer peak positions of $\epsilon/\omega$.}
\label{pic:prob}
\end{center}
\end{figure}

\begin{figure}[h!!!!!!!!!!!!!!!!!]
\begin{center}
\includegraphics[width=\linewidth]{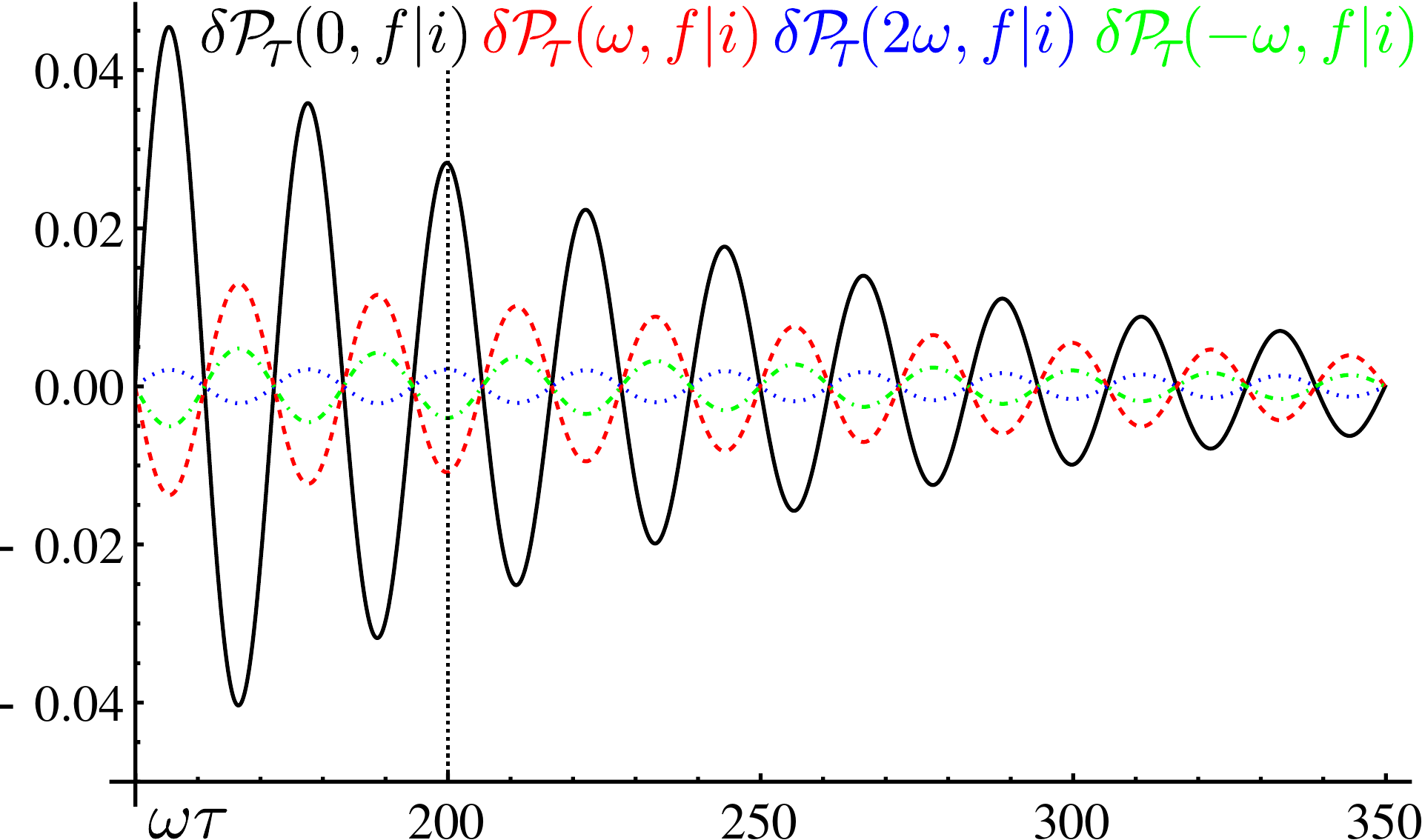}

\caption{(Color online) Quantum correction $\delta\mathcal{P}_\tau(\epsilon,f|i)$ for $f\neq i$ as a function of time for different values of $\epsilon=0$ (black), $\epsilon=\omega$ (red, dashed), $\epsilon=2\omega$ (blue, dotted) and $\epsilon=-\omega$ (green, dashed dotted). The intersection at $\tau=200/\omega$ denotes the time, where the conditional probabilities of \fig{pic:prob} have been calculated. }
\label{pic:deltaP}
\end{center}
\end{figure}

The conditional PDFs $\mathcal{P}_\tau(\epsilon, f|i)$ are depicted in \fig{pic:prob} and were calculated via numerical Fourier Transform of eq. \eqref{eq:GF}. The positions of the peaks are given by $n\omega +\sigma \Omega$, where 
$\sigma=0,+,-$ and $n$ is integer. As mentioned above these conditional probabilities contain considerable quantum 
contributions (see Fig.~\ref{pic:prob}), which we can calculate analytically. It turns out that these corrections appear only for $\epsilon = n\omega$, i.e., only for the central peaks in Fig.~\ref{pic:prob}. We obtain 
\begin{align}
 \delta\mathcal{P}_\tau(\epsilon,f|i)\!&= 
 -(-1)^{\langle f | i \rangle} \cos(\Omega \tau) \frac{\sin^2\theta}{2}\,  e^{-\tau\Gamma_\varphi^0} \nn\\
 &\times \sum_n \delta(\epsilon - n\omega)
 \left(\frac{i \gamma(\omega)^{1/2}}{ \gamma(-\omega)^{1/2}}\right)^n\!\! J_n\left[ i\eta \tau \right]
\end{align}
with $J_n[i\eta\tau]$ being the Bessel function of first kind. For the sake of legibility, we introduce the abbreviation 
$\eta\equiv\sin^2(\theta)(\gamma(\omega)\gamma(-\omega))^{1/2}/2$ whereas the dephasing rate $\Gamma_\varphi^0 \equiv (\gamma^+(\Omega)+\gamma^-(-\Omega))/2+\sin^2\theta(\gamma(\omega)  +\gamma(-\omega))/4$
is a part of (\ref{Gphi}) that does not contain the exponents of the counting field, i.e., $e^{\pm i\omega\lambda}$. The corrections are depicted in \fig{pic:deltaP} as a function of the driving time. As expected, the quantum part decays and oscillates with the frequency of the level splitting $\Omega$. The dotted vertical line indicates the time 
$\tau=200/\omega$ at which the PDFs in \fig{pic:prob} were calculated. 
We can easily show that our generating function obeys $\chi_\tau(\lambda,f|i) = \chi_{\tau}(-\lambda +i\beta,i|f)$, leading to the detailed fluctuation relation~\cite{Jarzynski-2000} 
\begin{align}\label{eq:FR}
 \frac{\mathcal{P}_\tau(\epsilon,f|i)}{\mathcal{P}_{\tau}(-\epsilon,i|f)}= e^{\epsilon \beta} \ .
\end{align}
Generally, we should have related the PDF $\mathcal{P}_{\tau}$ with the time reversed PDF 
$\mathcal{P}_{\tau,B}$, where not only the initial and the final states are interchanged, but also the driving protocol 
is time reversed~\cite{Campisi-2011, Esposito-2009}. In our case, however, the time inversion of the driving protocol amounts 
to a mere phase shift, which is immaterial within the RWA.

In conclusion, we have calculated the conditional probability densities of energy dissipated by a driven two-level system. 
In the non-trivial case of transversal coupling the energy exchanged between the system and the bath can take the values of  
multiples of the driving frequency $\omega$ shifted by $\pm $ the level splitting $\Omega$ in the rotating frame. 
We confirm the validity of the detailed fluctuation theorem by Jarzynski and consequently the Crooks relations in our system. 

The main result of our studies is the relatively large quantum corrections to the conditional probabilities, which oscillates with frequency $\Omega$ and decays on the time scale of the dephasing time. Observing these quantum corrections would constitute a first test of fluctuation relations in the quantum regime.

{\it Acknowledgements.} We thank J. Pekola and U. Briskot for valuable discussions. We acknowledge financial support of the German Science Foundation (DFG), the German-Israeli Foundation (GIF) and MEXT kakenhi ``Quantum Cybernetics" (No. 21102003).

\end{document}